\documentclass[12pt]{article}
\usepackage{amsmath,amssymb,epsfig,color}
\usepackage{psfrag}
\usepackage{graphicx}
\usepackage{bm}

\newcommand{\be}{\begin{equation}}  
\newcommand{\ee}{\end{equation}}

\begin{document}

\begin{titlepage}

\begin{flushright}
EFI Preprint 10-21\\
August 26, 2010
\end{flushright}

\vspace{0.7cm}
\begin{center}
\Large\bf 
Model independent extraction of 
the proton charge radius from electron scattering
\end{center}

\vspace{0.8cm}
\begin{center}
{\sc   Richard J. Hill and Gil Paz}\\
\vspace{0.4cm}
{\it Enrico Fermi Institute and Department of Physics \\
The University of Chicago, Chicago, Illinois, 60637, USA
}
\end{center}
\vspace{1.0cm}
\begin{abstract}
  \vspace{0.2cm}
  \noindent
Constraints from analyticity are combined with experimental 
electron-proton scattering data to determine the proton charge
radius.  
In contrast to previous determinations, we provide a systematic procedure
for analyzing arbitrary data without model-dependent assumptions on the form factor 
shape. 
We also investigate the impact of including electron-neutron 
scattering data, and $\pi\pi\to N\bar{N}$ data.
Using representative datasets we find 
$r_E^p=0.870 \pm 0.023 \pm 0.012 \,{\rm fm}$ using just proton scattering data;
$r_E^p=0.880^{+0.017}_{-0.020} \pm 0.007 \,{\rm fm}$ adding neutron data; and 
$r_E^p=0.871 \pm 0.009 \pm 0.002 \pm 0.002 \,{\rm fm}$ adding $\pi\pi$ data. 
The analysis can be readily extended to other nucleon form factors and derived 
observables. 
\end{abstract}
\vfil

\end{titlepage}

\section{Introduction}

The electromagnetic form factors of the nucleon 
provide basic inputs to precision tests of the Standard Model. 
In particular, the root mean square (RMS) proton charge radius as determined by 
the form factor slope\footnote{$G_E^p$ is defined in Section~\ref{sec:def}.} , 
\be
G_E^p(q^2) = 1 + {q^2\over 6} \langle r^2 \rangle_E^p + \dots, 
\ee
is an essential input to hydrogenic bound state
calculations~\cite{Udem:1997zz,Melnikov:1999xp}.  Recent experimental
results suggest a discrepancy between the charge radius inferred from the 
Lamb shift in muonic hydrogen~\cite{Pohl:2010zz},  
$r_E^p \equiv \sqrt{\langle r^2
\rangle_E^p }=0.84184(67)$~fm, and the CODATA value, 
$r_E^p =0.8768(69)$~fm, 
extracted mainly from (electronic) hydrogen spectroscopy \cite{Mohr:2008fa}. 
The charge radius can also be
extracted from elastic electron-proton scattering data. The 2010
edition of the Review of Particle Physics lists 12 such determinations
that span the range of 0.8-0.9~fm~\cite{{Nakamura:2010}}, most with 
quoted uncertainties of 0.01-0.02~fm. These determinations 
correspond to analyses of different datasets and different functional
forms of $G_E^p(q^2)$ that were fit to the data over a period of 50
years.

Extraction of the proton charge radius from scattering data is
complicated by the unknown functional behavior of the form factor.  We
are faced with the tradeoff between introducing too many parameters
(which limits predictive power) and too few parameters (which biases
the fits).  Here we describe a procedure that provides
model-independent constraints on the functional behavior of the form
factor.  The constraints make use of the known analytic properties of
the form factor, viewed as a function of the complex variable $t=q^2=-Q^2$.

\begin{figure}[h!]
\begin{center}
\psfrag{a}{$\!\!\!\!\!\!\!-Q^2_{\rm max}$} 
\psfrag{b}{$\!\!\!\!4m_\pi^2$} 
\psfrag{t}{$t$}
\psfrag{z}{$\!z$}
\epsfig{file=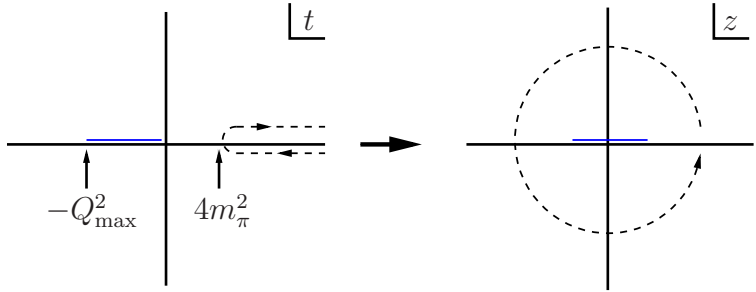,width=10cm}
\caption{\label{cutplane} Conformal mapping of the cut plane to the unit circle.}
\end{center}
\end{figure}  

\begin{figure}[h!]
\begin{center}
\psfrag{x}{$Q^2_{\rm max}$}
\psfrag{y}{$G_E^p$}
\psfrag{z}{$z$}
\psfrag{w}{$G_E^p$}
\epsfig{file=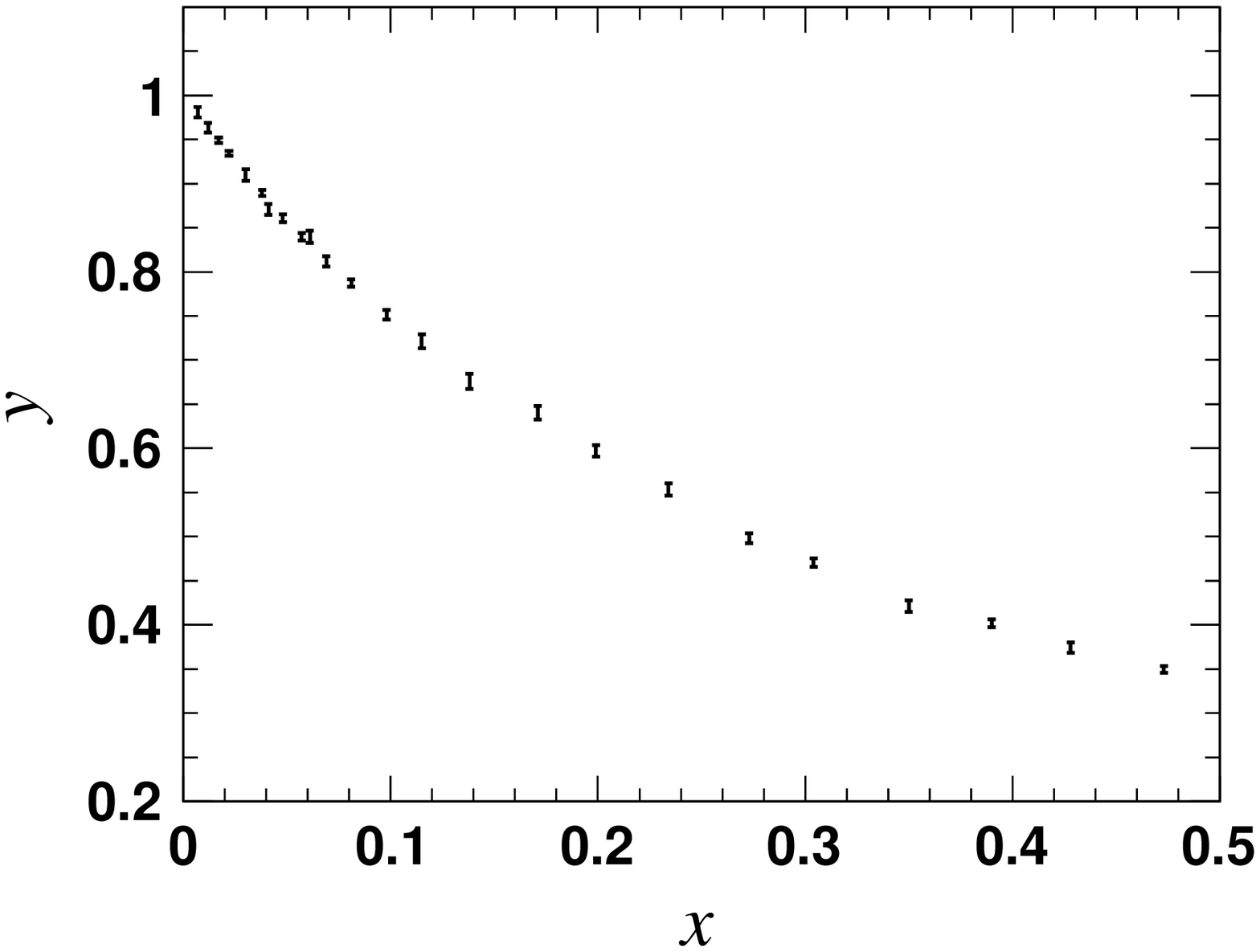,width=8cm}
\epsfig{file=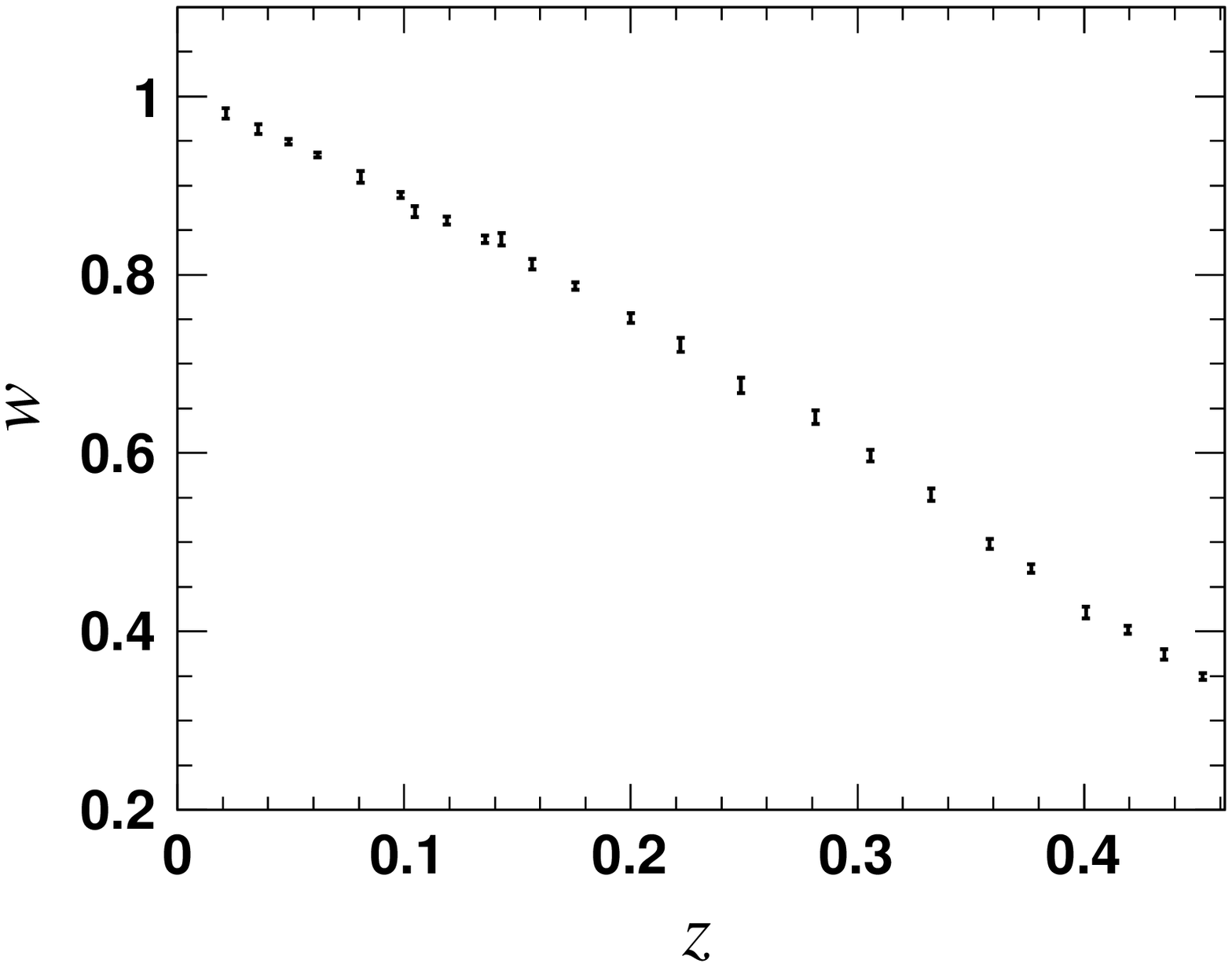,width=8cm}
\caption{\label{fig:qz} Form factor as a function of $Q^2$ and as a function of $z$.  
Here we choose $t_0=0$ in the definition of $z$, 
and plot data from \cite{Arrington:2007ux} 
for $0\le Q^2 \le 0.5\,{\rm GeV}^2$.   
}
\end{center}
\end{figure}

As illustrated in figure \ref{cutplane}, the form factor is analytic outside of a cut
at timelike values of $t$, \cite{Federbush:1958zz} beginning at the two-pion production threshold, 
$t\ge 4m_\pi^2$.%
\footnote{
Here and throughout, $m_\pi=140\,{\rm MeV}$ denotes the charged pion mass, 
and $m_N=940\,{\rm MeV}$ is the nucleon mass. 
}
In a restricted region of physical kinematics accessed 
experimentally, $-Q_{\rm max}^2 \le t \le 0$, the distance to singularities 
implies the existence of a small expansion parameter.   We begin by 
performing a conformal mapping of the domain of analyticity onto the unit circle: 
\be\label{eq:z}
z(t,t_{\rm cut},t_0) = {\sqrt{t_{\rm cut} - t} - \sqrt{t_{\rm cut} - t_0} \over \sqrt{t_{\rm cut} - t} + \sqrt{t_{\rm cut} - t_0}  } \,, 
\ee
where for this case $t_{\rm cut} = 4m_\pi^2$, and $t_0$ is a free parameter 
representing the point mapping onto $z=0$.   By the choice 
$t_0^{\rm opt} = t_{\rm cut} \left( 1 - \sqrt{1+ Q^2_{\rm max}/t_{\rm cut}} \right)$, the 
maximum value of $|z|$ is minimized: 
$|z| \le |z|_{\rm max} = [(1+Q_{\rm max}^2/t_{\rm cut})^\frac14 - 1]/[(1+Q_{\rm max}^2/t_{\rm cut})^\frac14 + 1]$. 
For example, with $Q^2_{\rm max} = 0.05\,{\rm GeV}^2$,  $0.5\,{\rm GeV}^2$, 
we find $|z|_{\rm max} = 0.062$, $0.25$.    Expanding the form factor as
\be \label{eq:zexpand}
G_E^p(q^2) = \sum_{k=0}^\infty a_k \, z(q^2)^k \,,
\ee
we find that the impact of higher order terms are suppressed by powers of
this small parameter%
\footnote{
Physical observables are independent of the choice of $t_0$, 
which can be viewed as the choice of an expansion ``scheme''.   
$|z|_{\rm max}$ defined in this way gives a convenient estimation of 
the impact of higher-order terms. 
}.
As we will see below, the coefficients multiplying $z^k$ 
are bounded in size, guaranteeing that a finite number of parameters are necessary to 
describe the form factor with a given precision.  
Figure~\ref{fig:qz} illustrates the manifestation of this fact in the 
form factor data.   As expected, the curvature is smaller in the $z$ variable than in 
the $Q^2$ variable.  

Expansions of the form (\ref{eq:z}) are a standard tool in analyzing 
meson transition form factors~\cite{Hill:2006ub, Bourrely:1980gp,
Boyd:1994tt,Boyd:1995sq,Lellouch:1995yv,Caprini:1997mu,Arnesen:2005ez,
Becher:2005bg,Hill:2006bq,Bourrely:2008za,Bharucha:2010im}.  
A complicating feature in the present application to nucleon form factors 
is the contribution of the subthreshold region $4m_\pi^2 \le t \le 4m_N^2$ in the relevant
dispersion integral.    

The rest of the paper is structured as follows. In Section
\ref{sec:fits} we demonstrate the application of the $z$ expansion in
some illustrative fits and compare it to other expansions that
appear in the literature. One of the main advantages of the $z$
expansion is that the expansion coefficients can be bounded using 
knowledge about ${\rm Im}G_E^p$ in the time-like region. In Section
\ref{sec:bounds} we discuss these bounds. In Section \ref{sec:more} we
discuss several possibilities of reducing the error on the charge
radius by including more experimental data, namely: high $Q^2$ data,
neutron scattering data, and $\pi\pi$ data. Finally, we discuss our
results in Section \ref{sec:discussion}.

\section{Illustrative fits \label{sec:fits}} 

Let us consider the six datasets tabulated by Rosenfelder~\cite{Rosenfelder:1999cd} 
(denoted in \cite{Rosenfelder:1999cd} as S1, S2, R, B1, B2, M)
This will allow us to compare in detail the
results of our fit to previous analyses.   For definiteness, we take all data 
points in \cite{Rosenfelder:1999cd} 
with corrections from magnetic form factor contributions $\Delta_{\rm mag}\le 0.15$. 
The resulting dataset has 85 points with $Q^2 \lesssim 0.04\,{\rm GeV}^2$. 

We will fit to three types of parameterization.  The first is 
a simple Taylor series expansion, 
\be\label{eq:taylor}
G_E^{p}(q^2) = 1 + a_1 {q^2 \over t_{\rm cut}} + a_2 \left(q^2 \over t_{\rm cut}\right)^2 + \dots \,, 
\ee
where we choose to work in units $t_{\rm cut}=4m_\pi^2$. 
The second is a continued fraction expansion put forward in \cite{Sick:2003gm}, 
\begin{align}\label{eq:cf}
G_E^{p}(q^2) &= {1 \over 1 + a_1 {q^2/t_{\rm cut}\over 1 + a_2 {q^2/t_{\rm cut}\over 1 + \dots}}} 
= 1 - a_1 {q^2\over t_{\rm cut}} + (a_1 a_2 + a_1^2) \left(q^2 \over t_{\rm cut}\right)^2 + \dots \,.
\end{align}
We are not aware of a motivation for this ansatz from first principles, but it
has been used to obtain one of the widely quoted values of the proton 
charge radius from electron scattering.   
The third is the $z$ expansion described in the Introduction,
\begin{align}
G_E^{p}(q^2) &= 1 + a_1 z(q^2) + a_2 z^2(q^2) + \dots 
= 1 -{a_1\over 4} {q^2\over t_{\rm cut}}  + \left( -{a_1\over 8} + {a_2 \over 16} \right) \left( q^2\over t_{\rm cut}\right)^2 + \dots 
\,,
\end{align}
where $z(q^2)=z(q^2,t_{\rm cut},t_0=0)$.  As explained below, the coefficients in this expansion
are bounded; for definiteness here we take $|a_k| \le 10$. 

\begin{table}
\begin{center}
\begin{tabular}{c|ccccc}
& $k_{\rm max}$ = 1 & 2 & 3 & 4 & 5 
\\
\hline 
polynomial & $836^{+8}_{-9}$ & $867^{+23}_{-24}$ & $866^{+52}_{-56}$ & $959^{+85}_{-93}$ & $1122^{+122}_{-137}$ 
\\
& $\chi^2$= 34.49 & 32.51 & 32.51 & 31.10 & 28.99 
\\
continued fraction & $882^{+10}_{-10}$ & $869^{+26}_{-25}$ & $-$ &$-$&$-$ 
\\
& $\chi^2$=32.81 & 32.51
\\
$z$ expansion (no bound) & $918^{+9}_{-9}$ & $868^{+28}_{-29}$ & $879^{+64}_{-69}$ & $1022^{+102}_{-114}$ & $1193^{+152}_{-174}$
\\
& $\chi^2$=36.14 & 32.52 & 32.48 & 30.35 & 28.92 
\\
$z$ expansion ($|a_k|\le 10$) & $918^{+9}_{-9}$ & $868^{+28}_{-29}$ & $879^{+38}_{-59}$ & $880^{+39}_{-61}$ & $880^{+39}_{-62}$ 
\\
& $\chi^2$=36.14 & 32.52 & 32.48 & 32.46 & 32.45 
\end{tabular} 
\end{center}
\caption{\label{tab:simplefits} 
Proton charge radius extracted from data of Table~1 of \cite{Rosenfelder:1999cd}  
($Q^2 \lesssim 0.04\,{\rm GeV}^2$) 
in units of $10^{-18}$~m, 
using different functional behaviors of the form factor.     
Dashes denote fits that do not constrain the slope to be positive. 
} 
\end{table}

We perform fits by minimizing a $\chi^2$ function, 
\be
\chi^2 = \sum_{i,\,j} ( {\rm data}_i - {\rm theory}_i ) E^{-1}_{ij} (  {\rm data}_j - {\rm theory}_j )  \,,
\ee
where the error matrix is formed by adding in quadrature the
quoted statistical errors, assumed uncorrelated, and  normalization error, assumed fully correlated within
each dataset.
In the notation of Table~1 of Ref.~\cite{Rosenfelder:1999cd} we use for each experiment, (note that $\delta_{\rm norm}$ refers
to the error in the {\it cross section})~\footnote{ 
We obtain similar results by floating the normalization of each experiment and
constraining the scale factors by an additional contribution to $\chi^2$ (as done in \cite{Rosenfelder:1999cd}) 
or by performing the fits at fixed (unit) normalization and assigning an additional error 
obtained by adding in quadrature the shift induced by redoing the fits with shifted normalization (as done in \cite{Sick:2003gm}). 
}
\be
E_{ij} = (\delta G_E)^2_i \delta_{ij} +  ( \delta_{\rm norm}/2 )^2 (G_E)_i (G_E)_j \,.
\ee
Errors for the form factor slope are computed by finding the $\Delta \chi^2=1$ range%
\footnote{
  We have performed these computations in both MAPLE and MATHEMATICA ,  and
have also checked our results using MINOS errors in MINUIT.   
}.

As can be seen from Table~\ref{tab:simplefits}, 
the fits with one free parameter differ by many standard deviations. 
Fits with two free parameters agree well, while fits with three or more parameters become increasingly 
unconstrained for the polynomial and continued fraction expansions,  
as well as for the $z$ expansion when no  constraints on the expansion coefficients are in place. 
In particular, for $k_{\rm max} \ge 3$ in the continued fraction 
expansion, no meaningful fit can be performed (e.g., the slope is not constrained to be positive). 

These results illustrate the problem to be addressed: without detailed knowledge of the 
functional behavior of the form factor, we risk using either too few parameters and 
biasing the fit; or too many parameters and losing predictive power.  
Note that performing trial fits on model data as in \cite{Sick:2003gm} is 
also problematic; some assumption must be made on the functional behavior of the form 
factor in creating the model datasets.
To make model independent statements requires identifying 
a bounded class of functions that is guaranteed  to contain the true form factor, 
yet is sufficiently restrictive to retain predictive power.  
The following section describes such a class of functions. 

\section{Dispersive bounds\label{sec:bounds}} 

The above fit to the $z$ expansion with a bound on the coefficients illustrates our
basic methodology. The present section justifies the $|a_k|\le 10$ bound, and demonstrates how
further constraints can be obtained by disentangling the isoscalar and isovector components of
the form factor. 

\subsection {Form factor definitions \label{sec:def}}

For completeness we list definitions of the various form
factors. The Dirac and Pauli form factors, $F^N_1$ and $F^N_2$, 
respectively, are defined by \cite{Foldy:1952,Salzman:1955zz}
\be
\langle N(p')|J_\mu^{\rm em}|N(p)\rangle=\bar u(p')
\left[\gamma_\mu F^N_1(q^2)+\frac{i\sigma_{\mu\nu}}{2m_N}F^N_2(q^2)q^\nu\right]u(p)\,,
\ee
where $q^2=(p'-p)^2=t$ and $N$ stands for $p$ or $n$. 
The Sachs electric and magnetic form factors
are related to the Dirac-Pauli basis by \cite{Ernst:1960zza}
\begin{align}
G^N_E(t) = F^N_1(t) + {t\over 4m_N^2} F^N_2(t) \,, \quad
G^N_M(t) = F^N_1(t) + F^N_2(t) \,. 
\end{align}
At $t=0$ they are~\cite{Nakamura:2010}
$G^p_E(0) = 1$,
$G^n_E(0) = 0$,
$G^p_M(0) = \mu_p \approx 2.793$, 
$G^n_M(0) = \mu_n\approx -1.913$.
We write the isoscalar and isovector form factors as
\begin{align}
G_E^{(0)} = G_E^{p} + G_E^{n} \,, \quad
G_E^{(1)} = G_E^{p} - G_E^{n} \,, 
\end{align} 
such that at $t=0$ they are,
$G_E^{(0)}(0) = 1$, 
$G_E^{(1)}(0) = 1$,  
$G_M^{(0)}(0) = \mu_p+\mu_n$ 
$G_M^{(1)}(0) = \mu_p-\mu_n$.
Notice that $G^{(0)}_{E,M}=2G^{s}_{E,M}$, $G^{(1)}_{E,M}=2G^{v}_{E,M}$ for $G^{s,v}_{E,M}$ of \cite{Belushkin:2006qa}.

\subsection{Dispersive bounds}

The analytic structure in the $t$ plane illustrated in Fig.~\ref{cutplane} 
implies the dispersion relation, 
\begin{align}
G_E^p(t) = {1\over \pi} \int_{t_{\rm cut}}^\infty {dt^\prime}\, {{\rm Im}G_E^p(t^\prime + i0) \over t^\prime - t } \,. 
\end{align}
Knowledge of ${\rm Im}G_E^p$ over the cut translates into information about the
coefficients in the $z$ expansion. We begin with a general discussion of these
relations.

Let us consider a general function with the analytic structure as in Fig.~\ref{cutplane},
$G(t)=\sum_{k=0}^\infty a_k z(t)^k$. Equation~(\ref{eq:z}) maps points
just above (below) the cut in the $t$ plane onto points in the lower
(upper) half unit circle in the $z$ plane. Parameterizing the unit
circle by $z(t)=e^{i\theta}$ and solving (\ref{eq:z}) for $t$, we find
\be
t=t_0+\frac{2(t_{\rm cut}-t_0)}{1-\cos \theta}\equiv t(\theta)\,. 
\ee  
We can now use the orthogonality of $z^k$ over the unit circle to find 
\be
a_k=\frac{1}{\pi}\int_0^\pi d\theta\,{\rm Re}\, G[ t(\theta)+i0]\,\cos(k\theta)-
\frac{1}{\pi}\int_0^\pi d\theta\,  {\rm Im}\, G[t(\theta)+i0 ] \,\sin(k\theta) \,.
\ee
Since $G$ is analytic, $a_k=0$ for $k<0$, and therefore 
\begin{align}\label{eq:fourier} 
a_0&=\frac{1}{\pi}\int_0^\pi d\theta\,{\rm Re}\, G[ t(\theta)+i0]=G(t_0) \,,
\nonumber\\
a_k&=-\frac{2}{\pi}\int_0^\pi d\theta\,  {\rm Im}\, G[t(\theta)+i0 ] \,\sin(k\theta) 
= {2\over \pi} \int_{t_{\rm cut}}^\infty {dt\over t-t_0} \sqrt{ t_{\rm cut} - t_0 \over t - t_{\rm cut}} 
{\rm Im}G(t)
\sin[ k\theta(t) ] 
\,,\quad k\ge 1\,.
\end{align}

The coefficients in the expansion (\ref{eq:zexpand}) can also be used to 
construct a norm of the form factor in the mathematical sense.   
To keep the discussion general, let us introduce a function $\phi$ 
sharing the domain of analyticity of $G$, and write 
\be\label{eq:phizexpand}
 \phi G = \sum_{k=0}^\infty a_k z^k \,.
\ee
Consider the class of norms specified by 
\be
||\phi G||_{p} = \left( \sum_k |a_k|^{p} \right)^{1\over p} \,. 
\ee
In particular, the ``uniform norm'' is equal to the maximum coefficient size,
$|| \phi G ||_\infty = \sup_{k} {|a_k|}  = \lim_{p\to \infty} ||\phi G||_{p}$. 
The case $p=2$ is  of special interest since the norm is easily related to a dispersion integral, 
\be\label{eq:norm2}
||\phi G||_{2} = \left( \sum_k a_k^2 \right)^\frac12 = \left( \oint {dz\over z} |\phi G|^2 \right)^\frac12
= \left( 
{1\over \pi} \int_{t_{\rm cut}}^\infty {dt \over t-t_0} \sqrt{ t_{\rm cut} - t_0 \over t - t_{\rm cut} } |\phi G|^2 
\right)^\frac12 
\,.
\ee
The finiteness of $||\phi G||_2$ shows that the coefficients $a_k$ are not only bounded, 
but must {\it decrease} in size for sufficiently large $k$.   
The relation $||\phi G||_{\infty} \le ||\phi G||_2$ indicates that $||\phi G||_2$ may overestimate the 
actual size of the relevant coefficients in certain cases.  We proceed to consider a vector 
dominance model to illustrate this feature and then turn to a more detailed analysis 
of the spectral functions.  

\subsection{Vector dominance ansatz} 

\begin{table}
\begin{center}
\begin{tabular}{cc|cccc}
& & $t_0=0$ & $t_0=t_0^{\rm opt}(0.5\,{\rm GeV}^2)$ 
\\
\hline 
$\phi=1$ 
& $||G^{(0)}_E||_2/G^{(0)}_E(t_0)$ & 7.6 & 12.1  
\\
& $||G^{(1)}_E||_2/G^{(1)}_E(t_0)$ & 2.5 & 3.9 
\\
$\phi=\phi_{\rm OPE}$
& $||\phi^{(0)}G^{(0)}_E||_2/\phi^{(0)}(t_0) G^{(0)}_E(t_0)$ & 14.4 & 23.5 
\\
& $||\phi^{(1)}G^{(1)}_E||_2/\phi^{(1)}(t_0) G^{(1)}_E(t_0)$ & 4.6 & 6.7 
\\\hline
$\phi=1$ 
&
$\left. 2\sqrt{ t_{\rm cut}-t_0 \over m_V^2-t_{\rm cut} } \right|_{I=0}$ 
& 1.3 & 1.8 
\\
&
$\left. 2\sqrt{ t_{\rm cut}-t_0 \over m_V^2-t_{\rm cut} } \right|_{I=1}$ 
& 0.78 & 1.3
\end{tabular} 
\end{center}
\caption{\label{tab:norm}
Typical bounds on the coefficient ratios $ \sqrt{ \sum_k {a_k^2/a_0^2} }$ (upper part of table) 
and $|a_k/a_0|$ (lower part) in a vector dominance ansatz.  $\phi_{\rm OPE}$  
is defined in Eq.(\ref{eq:phiunitary}).
} 
\end{table}

In many applications, the $||\cdot ||_2$ norm is used in conjunction with ``unitarity bounds''
obtained by identifying the dispersive integral with a physical production rate.    
In the present example, dominant contributions to the integral arise from the region below the 
two-nucleon production threshold, and we must turn to different methods of analysis.    
For example, employing a vector dominance ansatz in the appropriate channel, 
Table~\ref{tab:norm} displays estimates for the quantity 
$|| \phi G ||_2/\phi(t_0)G(t_0) = \sqrt{ \sum_k {a_k^2 / a_0^2} }$, 
for different choices of the functional form of $\phi$ and the value of 
$t_0$~\footnote{
For this purpose we estimate $G_E(t_0)$ using a dipole 
ansatz for the form factor, $G_{E}(t) \sim 1/(1-t/0.71\,{\rm GeV}^2)^2$.
}.
The effects of the leading resonance in each channel are represented by a Breit Wigner profile~\cite{Hohler:1974ht}, 
\be
F_i^{(I=0)} \sim {\alpha_i m_\omega^2 \over m_\omega^2 - t - i\Gamma_\omega m_\omega} \,,
\quad
F_i^{(I=1)} \sim {\beta_i m_\rho^2 \over m_\rho^2 - t - i\Gamma_\rho m_\rho} \,, 
\ee
with
$\alpha_1 \approx 1$, $\alpha_2\approx -0.12$, 
$m_\omega=783\,{\rm MeV}$, $\Gamma_\omega = 8.5\,{\rm MeV}$ 
for the isoscalar channel; and
 $\beta_1\approx 1$, $\beta_2\approx 3.7$, 
$m_\rho=775\,{\rm MeV}$, $\Gamma_\rho = 149\,{\rm MeV}$ 
for the isovector channel. At $\Gamma=0$, the ansatz is normalized to 
the $t=0$ values in Section~\ref{sec:def}. 

We note that in the isoscalar case, the rather large size of the estimated norm is due to the 
narrow width of the $\omega$ resonance; in fact, in the limit of an infinitely narrow resonance, 
the quantity $||G||_2$ diverges, as seen from (\ref{eq:norm2}).   
Closer examination indicates that the large norm is due not to the coefficients growing in size, but 
rather to a sequence of coefficients whose slow fall-off causes
a slow convergence for the sum $\sum_k a_k^2$.    
A straightforward computation shows that the expansion coefficients for an infinitely
narrow pole, $G(t) = G(0)/(1-t/m_V^2)$, are for $k\ge 1$, 
\be\label{eq:narrow}
{a_k \over a_0} =  {-2}\sqrt{t_{\rm cut} -t_0 \over m_V^2 - t_{\rm cut}} \sin \left[ 
2k \arcsin\left( \sqrt{t_{\rm cut}-t_0\over m_V^2-t_0} \right) \right]  \,. 
\ee
In particular, $|a_k/a_0| \le 2\sqrt{ (t_{\rm cut}-t_0)/(m_V^2-t_{\rm cut}) }$.   
This approximation to the uniform norm is also displayed in Table~\ref{tab:norm}. 
 
Equations (\ref{eq:fourier}) and 
 (\ref{eq:norm2})   are model-independent, whereas the 
approximations based on  the vector dominance ansatz employed in Table~\ref{tab:norm} 
are model dependent. 
This ansatz aims simply to capture the order of 
magnitude of the coefficients, which is sufficient in practice to constrain the form 
factor fits.  
The conclusion is that $|a_k|\le 10$ is a very conservative estimate for this ansatz. 

\subsection{Explicit $\pi\pi$ continuum \label{sec:pipi}}

We can be more explicit in the case of the isovector form factor
expansion, where the leading singularities are due to $\pi\pi$
continuum contributions that are in principle constrained by measured $\pi\pi$
production and $\pi\pi \to N\bar{N}$ annihilation rates
\cite{Federbush:1958zz, Frazer:1960zzb, Belushkin:2006qa}:
\begin{align}\label{eq:pipiNN}
{\rm Im}\, G_E^{(1)}(t) = {2\over m_N \sqrt{t}} \left(t/4-m_\pi^2 \right)^\frac32 F_\pi(t)^* f^{1}_+(t) \,,
\end{align}
where $F_\pi(t)$ is the pion form factor (normalized according to $F_\pi(0)=1$) and
$f^1_+(t)$ is a partial amplitude for $\pi\pi\to N\bar{N}$.    Using that these quantities 
share the same phase~\cite{Frazer:1960zzb}, we may substitute absolute values.  
Strictly speaking, this relation holds up to the four-pion threshold, $t\le 16 m_\pi^2$. 
For the purposes of estimating coefficient bounds, 
we will take the extension of (\ref{eq:pipiNN}) assuming phase equality through 
the $\rho$ peak as a model for the total $\pi\pi$ continuum contribution.  

For $|F_\pi(t)|$ we take an interpolation using 
the four $t$ values close to production threshold from \cite{Amendolia:1983di} 
($0.101$ to $0.178\,{\rm GeV}^2$), and 43 $t$ values from \cite{Achasov:2005rg} 
($0.185$ to $0.94\,{\rm GeV}^2$).   
Values for $f_+^1(t)$ are taken from Table~2.4.6.1 of \cite{Hohler}.  
Evaluating (\ref{eq:fourier}) using (\ref{eq:pipiNN}) 
and the experimental data up to $t=0.8\,{\rm GeV}^2\approx 40\,m_\pi^2$
yields for the first few coefficients, 
at $\phi=1$ and $t_0=0$: 
$a_0 \approx 2.1$
$a_1 \approx -1.4$, $a_2 \approx -1.6$, $a_3\approx -0.9$, $a_4 \approx 0.2$.   
Using $|\sin(k\theta)|\le 1$ in the integral gives $|a_k| \lesssim 2.0$ for $k\ge 1$.    

The leading singularities in the isoscalar channel could in principle be analyzed using 
data for the $3\pi$ continuum.    Since we do not attempt to raise the isoscalar
threshold in our analysis, we content ourselves with a simple vector dominance model to estimate the 
coefficient bounds.  
The first few coefficients for the isoscalar 
form factor using (\ref{eq:narrow}) for a narrow $\omega$ resonance are: 
$a_0 =1$, $a_1 \approx -1.2$, $a_2\approx -0.96$, $a_3\approx 0.4$, $a_4 \approx 1.3$.  
We will compare the above values to those extracted from electron scattering data later. 
For the moment we note that a bound $|a_k|\le 10$ is conservative.  

\subsection{Choice of $\phi$} 

Let us return to the choice of $\phi$.   
We will consider three essentially different choices.  First, $\phi(t)=1$ is our default choice. 
We noted that for $\phi=1$ the dominant contributions to $||\phi G||_2$ are from narrow resonances.
We could negate the large contribution of the 
leading resonances by using for $\phi$ the inverse of a vector meson dominance (VMD) form factor.   
As a second choice, consider 
\begin{align}\label{eq:phivmd}
\phi_{\rm VMD}(t) = (m_V^2-t)/m_V^2 \,, 
\end{align}
where $m_V$ is the mass of the leading resonance in the appropriate channel, i.e., 
$\rho(770)$ for the isovector, $\omega(780)$ for the  isoscalar.  
Note that using $G_{E} \sim 1/t^2$ at large $t$, the dispersion integral remains convergent. 
There is no loss of model-independence here, since corrections to vector dominance are accounted for 
in the coefficients $a_k$.   
As discussed in Section~\ref{sec:timelike}, a third choice of $\phi$ is motivated by unitarity and an operator 
product expansion (OPE): 
\begin{align}\label{eq:phiunitary}
\phi_{\rm OPE}(t) &= {m_N \over \sqrt{6\pi}} { (t_{\rm cut} - t)^\frac14 \over (t_{\rm cut} - t_0)^\frac14 } \left[ z(t,t_{\rm cut},0)\over -t \right]^\frac14 
\left[ z(t,t_{\rm cut},t_0)\over t_0-t\right]^{-\frac12} \left[ z(t,t_{\rm cut},-Q_{\rm OPE}^2)\over -Q_{\rm OPE}^2-t \right]^\frac32 (4m_N^2 - t)^\frac14 \,,
\end{align}
where $t_{\rm cut}$ is appropriate to the chosen isospin channel.  
For definiteness, we choose $Q_{\rm OPE}^2=1\,{\rm GeV}^2$ in the unitarity-inspired $\phi$.    
In our final fits, we focus on $\phi=1$ and $t_0=0$ but demonstrate that the results 
are essentially unchanged for different choices. 

\subsection{Bounds on the region $t\ge 4m_N^2$ \label{sec:timelike}}

The contribution of the physical region $t\ge 4m_N^2$ to $||\phi G_E||_2$ is
\begin{align}
\delta ||\phi G_E||_2^2 = {1\over \pi} \int_{4m_N^2}^\infty {dt\over t-t_0} \sqrt{ t_{\rm cut} - t_0 \over t - t_{\rm cut}} |\phi G_E|^2 \,.
\end{align}
The cross section for $e^+ e^- \to N\bar{N}$ is~\cite{Cabibbo:1961sz}
\begin{align}
\sigma(t) = {4\pi\alpha^2\over 3 t} \sqrt{ 1 - {4m_N^2\over t}} \left( |G_M(t)|^2 + {2m_N^2\over  t} |G_E(t)|^2 \right) \,,
\end{align}
and thus for the proton electric form factor we have
\begin{align}\label{eq:dnorm}
\delta ||\phi G_E^p||_2^2 
= {1\over \pi} \int_{4m_N^2}^\infty {dt\over t-t_0} \sqrt{ t_{\rm cut} - t_0 \over t - t_{\rm cut}} 
|\phi|^2 
\left[ 
{ \sigma(t) \over \sigma_0(t) v(t) } {1\over |G_M/G_E|^2 + {2m_N^2/t}  }
\right]
\,,
\end{align}
where $\sigma_0 = 4\pi\alpha^2/3t$ and $v(t) = \sqrt{1-4m_N^2/t}$ is the nucleon velocity in the center-of-mass frame. 
Using the data from \cite{Ablikim:2005nn} (see also \cite{Pedlar:2005sj,Aubert:2005cb}), 
we can perform the integral from $t=4.0\,{\rm GeV}^2$ to $9.4\,{\rm GeV}^2$ assuming $|G_M^p/G_E^p|\lesssim 1$.%
\footnote{
For $|G_M/G_E|\ge 1$, the quantity in square brackets in (\ref{eq:dnorm}) is 
bounded by the quantity denoted by $|G|^2$ in \cite{Ablikim:2005nn}.   This inequality is 
satisfied experimentally in the $t$ range of interest. 
}
At $t_0=0$ and $\phi=1$, we find the result $\delta ||G_E^p||^2_2 \lesssim (0.03)^2$, to be added 
to the contribution from $t\le 4m_N^2$.    This result is obtained by using for $\sigma(t)$ 
the measured central value plus $1\sigma$ error.   The remaining integral above $t=9.4\,{\rm GeV}^2$ can 
be conservatively estimated by assuming a constant form factor beyond this point, yielding an 
additional $\delta ||G_E^p||_2^2 \approx (0.008)^2$.   
The neutron form factor can be treated similarly using the data from \cite{Antonelli:1998fv} for 
$t=3.61$ to $5.95\,{\rm GeV}^2$.    This leads to $\delta ||G_E^n||^2_2 \approx (0.05)^2$.  
The remainder at high $t$ assuming a constant form factor yields an additional $\delta ||G_E^n||_2^2 \approx (0.05)^2$. 
Similarly, using $|{\rm Im}G_E \sin k\theta|\le |G_E|$ the contribution of the 
timelike region to (\ref{eq:fourier}) is small: 
$|\delta a_k| \lesssim 0.011 + 0.004$ for the proton, and $|\delta a_k| \lesssim 0.013 + 0.025$ 
for the neutron.  We conclude that when estimating the bounds on coefficients, the physical timelike 
region can be safely neglected.   

Let us mention that we can bound the contribution of the physical timelike region by a perturbative 
quark-level computation.   Decompose the electromagnetic current correlation function as 
\begin{align}
\Pi^{\mu\nu}(q) &=  i \int d^4x\, e^{iq\cdot x} \langle 0 | T\{ J_{\rm em}^\mu(x), J_{\rm em}^\nu(0) \} |0\rangle  
= (q^\mu q^\nu - q^2 g^{\mu\nu} ) \Pi(q^2) \,.
\end{align} 
and define 
\begin{align}
\chi(Q_{\rm OPE}^2) = \left. \frac12 {\partial^2 \over \partial (q^2)^2} ( q^2 \Pi(q^2) ) \right|_{q^2=-Q^2_{\rm OPE}}  
= {1\over \pi} \int_{t_0}^\infty dt\, {t {\rm Im} \Pi(t) \over (t+Q_{\rm OPE}^2)^3 } \,.
\end{align}
The two-nucleon contribution to the correlator satisfies 
\begin{align} 
{\rm Im} \Pi(t)  
\ge 
{m_N^2 \over  6 \pi t} \sqrt{1-{4m_N^2\over t}}  |\phi G_E|^2  \,, 
\end{align} 
and hence with $\phi G_E = \sum_k a_k z^k$ and the choice of $\phi$ in (\ref{eq:phiunitary}), 
 \begin{align}
\chi(Q_{\rm OPE}^2)
\ge {1\over \pi} 
\int_{4m_N^2}^\infty {dt\over t-t_0}  
  \sqrt{ t_{\rm cut} - t_0 \over t-t_{\rm cut}} |\phi G_E|^2 
 \ge \delta  || \phi G_E ||_2^2  \,. 
 \end{align}
If we choose $Q_{\rm OPE}^2$ large enough, the function $\chi(Q_{\rm OPE}^2)$ is perturbatively calculable
as an operator product expansion: 
$\chi \approx \sum_f e_f^2 / 8\pi^2 Q_{\rm OPE}^2$ at leading order, where $e_f$ denotes 
the electric charge of a given quark flavor.
Choosing for illustration $Q_{\rm OPE}^2 = 1\,{\rm GeV}^2$, $n_f = 3$ light quark flavors,  
and $t_{\rm cut}=4m_\pi^2$, 
we find the bounds $\delta( \sum_k a_k^2 ) \sim (1.0)^2$ for $t_0=0$ 
and $\delta( \sum_k a_k^2/a_0^2 ) \sim (1.4)^2$ for $t_0=t_0^{\rm opt}(0.5\,{\rm GeV}^2)$.
We note that these ``unitarity bounds'' overestimate the contribution 
from the physical region $t\ge 4m_N^2$, due both to subthreshold 
resonance production, and to other channels, e.g., $N\bar{N}$ plus pions, above threshold.   
For this reason, we do not dwell on a more precise analysis of this bound, 
or on a separation into definite isospin channels.

\section{Proton charge radius extraction \label{sec:more}} 

We consider several possibilities to reduce the error bars for the proton 
charge radius extracted in Section~\ref{sec:fits}.   We first consider the inclusion 
of higher-$Q^2$ data. 
We then optimize the charge radius extraction by separating 
isoscalar and isovector components, recognizing that the isoscalar threshold is at $9m_\pi^2$. 
At the same time, we illustrate the (small) effect of different expansion schemes.
Finally, we consider the possibility to effectively raise 
the isovector threshold by constraining the spectral function between $4 m_\pi^2$ and $16 m_\pi^2$. 

\subsection{Including higher $Q^2$ data} 

\begin{figure}[ht]
\begin{center}
\psfrag{x}{$Q^2_{\rm max}$}
\psfrag{y}{$r_E^p ({\rm fm})$}
\epsfig{file=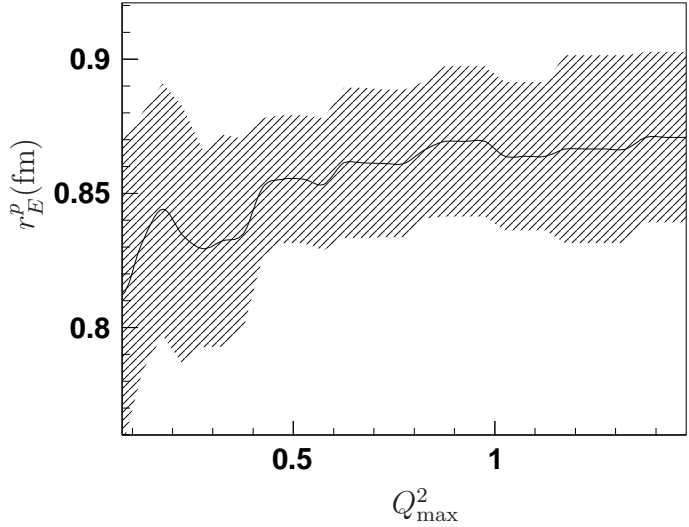,width=10cm}
\caption{\label{fig:Q2} Variation of the fitted proton charge radius as a function of 
maximum $Q^2$.  Fits of the proton data were performed with $k_{\rm max}=10$, $\phi=1$, 
$t_0=0$, $|a_k|\le 10$.   
Data from \cite{Arrington:2007ux}.
}
\end{center}
\end{figure}

We have argued that, taking the data tabulated in 
\cite{Rosenfelder:1999cd} at face value, the 
final entry in Table~\ref{tab:simplefits} is a model-independent determination of the 
proton charge radius: $r_E^p =  0.878^{+0.039}_{-0.062}\,{\rm fm}$. 
In the absence of further model-independent constraints on the form factors, 
obtaining a proton charge radius with smaller error requires further 
experimental input.   Here we investigate the impact of higher-$Q^2$ proton 
scattering data.  

Figure~\ref{fig:Q2} shows the central value and $1\sigma$ ($\Delta \chi^2=1$) 
error band obtained by fitting the electron-proton scattering 
data compiled by Arrington et al.~\cite{Arrington:2007ux}.  
We take $\phi=1$ and $t_0=0$, and include as many coefficients $a_k$ as necessary 
for the fits to stabilize.  
As the figure illustrates, for $Q^2\gtrsim {\rm few} \times 0.1\,{\rm GeV}^2$ the
impact of additional data is minimal.   While an ever greater number of 
coefficients $a_k$ at higher $k$ must be included to obtain convergence, the 
total error on the slope at $Q^2=0$ is not reduced.   
For later use, we note that the coefficients $a_{k=1,2,3}$ 
extracted from the fit at $Q^2_{\rm max}=1\,{\rm GeV}^2$ are
$-1.01(6)$, $-1.4^{+1.1}_{-0.7}$, $2^{+2}_{-6}$.   

\subsection{Raising the isoscalar threshold: inclusion of neutron data \label{sec:neutron}}

\begin{table}
\begin{center}
\begin{tabular}{l|ccccc}
& $k_{\rm max}=2$ & 3 & 4 & 5 & 6
\\
\hline
$\phi=1$, $t_0=0$, $|a_k|\le 10$ & $888^{+5}_{-5}$ & $865^{+11}_{-11}$ & $888^{+17}_{-22}$ & $882^{+21}_{-22}$ & $878^{+20}_{-19}$ 
\\
& $\chi^2=33.67$ & 23.65 & 21.80 & 21.13 & 20.47 
\\
$\phi=1$, $t_0=0$, $|a_k|\le 5$ & $888^{+5}_{-5}$ & $865^{+11}_{-11}$ & $881^{+10}_{-16}$ & $885^{+16}_{-21}$ & $882^{+18}_{-20}$
\\
& $\chi^2=33.67$ & 23.65 & 21.95 & 21.46 & 21.06 
\\
$\phi=\phi_{\rm VMD}$, $t_0=0$,  $|a_k|\le 10$ & $865^{+6}_{-6}$ & $874^{+12}_{-13}$ & $884^{+23}_{-24}$ & $879^{+24}_{+22}$ & $877^{+22}_{-20}$ 
\\
& $\chi^2=23.26$ & 22.50 & 22.15 & 21.59 & 21.09 
\\
\hline
$\phi=1$, $t_0=0$ & $888^{+5}_{-5}$ & $865^{+11}_{-11}$ & $880^{+13}_{-16}$ & $882^{+14}_{-18}$ & $882^{+15}_{-18}$
\\
& $\chi^2=33.67$ & 23.65 & 22.07 & 21.45 & 21.18
\\
$\phi=\phi_{\rm OPE}$, $t_0=0$ & $904^{+5}_{-5}$ & $861^{+10}_{-11}$ & $888^{+14}_{-21}$ & $883^{+20}_{-20}$
& $881^{+20}_{-19}$ 
\\
& $\chi^2=61.34$ & 24.38 & 21.62 & 20.86 & 20.51
\\
$\phi=\phi_{\rm OPE}$, $t_0=t_0^{\rm opt}(0.5\,{\rm GeV}^2)$ & 
$912^{+5}_{-5}$ & $869^{+9}_{-9}$ & $887^{+18}_{-19}$ & $881^{+20}_{-19}$ & $880^{+20}_{-19}$ 
\\
& $\chi^2=93.69$ & 22.54 & 21.05 & 20.32 & 20.32
\end{tabular} 
\end{center}
\caption{
RMS charge radius extracted using electron-proton and electron-neutron 
scattering data, and different schemes presented in the text.   The neutron 
form factor slope is constrained using (\ref{eq:ren}).   
A cut $Q^2_{\rm max}=0.5\,{\rm GeV}^2$ is enforced.   
In the lower part of the table, 
the bounds on $\sum_k a_k^2$ from Table~\ref{tab:norm} are multiplied by $4$. 
$\phi_{\rm VMD}$ and $\phi_{\rm OPE}$  
are defined in Eqs.(\ref{eq:phivmd}),(\ref{eq:phiunitary}).
\label{tab:schemes}
} 
\end{table}

We can separate the 
isoscalar from the isovector form factor, making use of the fact that the isoscalar cut 
is further away from $t=0$ than the isovector cut, translating to a smaller value of $|z|_{\rm max}$ as discussed in 
the Introduction.  
A combined fit of proton and neutron data can then be performed.  
For the proton form factor we again use the data from \cite{Arrington:2007ux}.  
For the neutron electric form factor, we use 20 data points from 
\cite{
Bermuth:2003qh, 
Eden:1994ji,
:2008ha, 
Glazier:2004ny,
Golak:2000nt,
Herberg:1999ud, 
Ostrick:1999xa,
Passchier:1999cj,
Plaster:2005cx,
Rohe:1999sh,
Warren:2003ma,
Zhu:2001md}.
We take as additional input the neutron charge radius from 
neutron-electron scattering length measurements~\cite{Nakamura:2010}:
\be\label{eq:ren}
\langle r^2\rangle_E^n = -0.1161(22) \, {\rm fm}^2 \,. 
\ee
Table~\ref{tab:schemes} shows the effect of different expansion schemes (choices
of $\phi$ and $t_0$) and coefficient bounds on the form factor slope determination. 
For later use, the coefficients $a_{k=1,2,3}$  
extracted from the fit for $Q^2_{\rm max}=1\,{\rm GeV}^2$, $\phi=1$, $t_0=0$ and $k_{\rm max}=8$ 
are 
$-1.99^{+0.13}_{-0.12}$, $0.3^{+1.5}_{-1.9}$, $-2^{+9}_{-6}$ for the isoscalar channel; 
and
$-1.20^{+0.06}_{-0.05}$, $-0.6^{+1.3}_{-1.2}$, $-2^{+6}_{-7}$ for the isovector channel. 
The sign and approximate magnitude of the first coefficients agree with  
the $\pi\pi$ continuum model, and the narrow-width $\omega$ resonance model 
mentioned in Section~\ref{sec:pipi}. 

\subsection{Raising the isovector threshold: inclusion of $\pi\pi$ data}

\begin{figure}[h!]
\begin{center}
\psfrag{x}{$Q^2_{\rm max}$}
\psfrag{y}{$r_E^p({\rm fm})$}
\epsfig{file=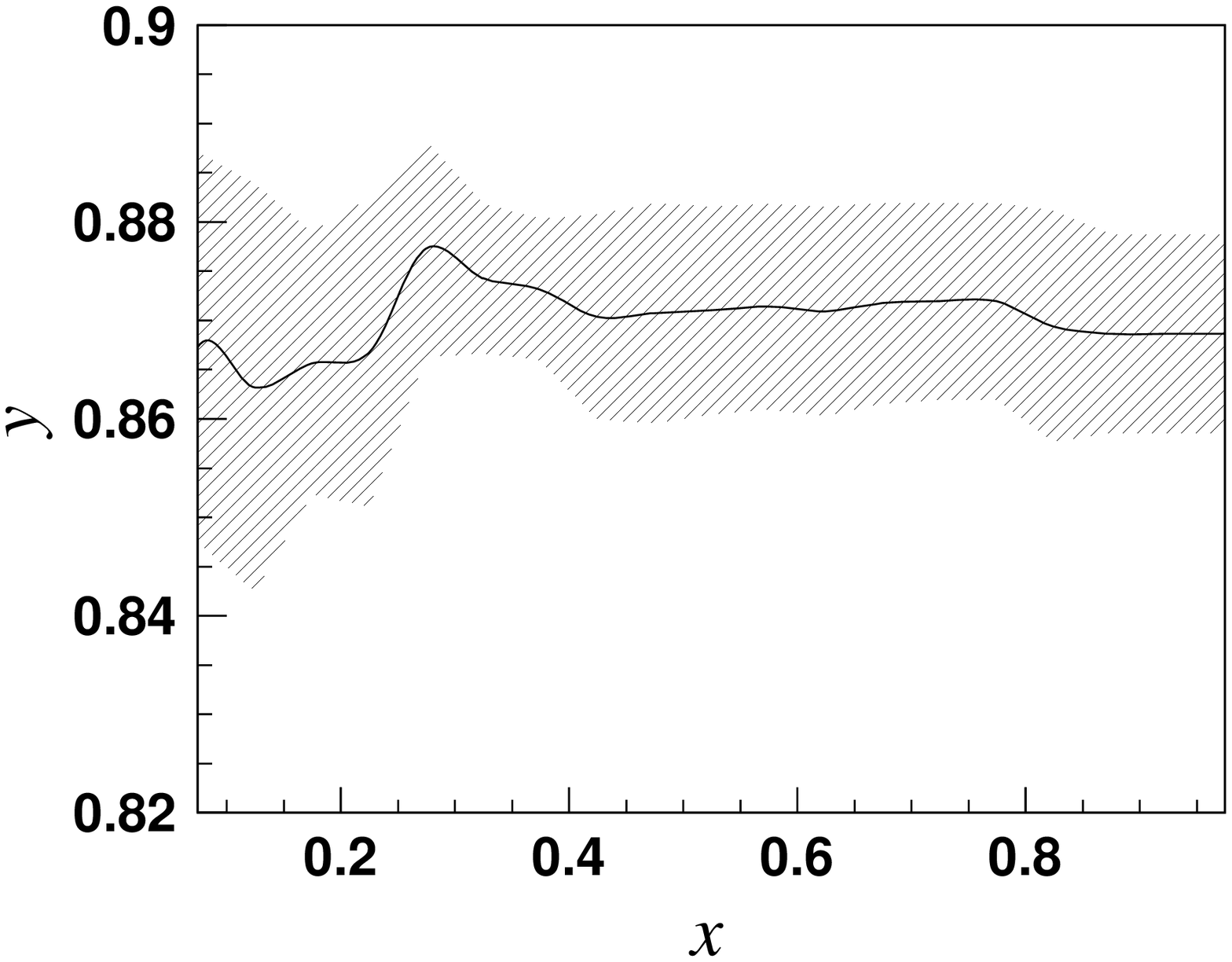,width=10cm}
\caption{Variation of the fitted proton charge radius as a function of 
maximum $Q^2$.  Fits were performed including proton data, neutron data and the $\pi\pi$ continuum contribution
to the isovector spectral function, as detailed in the text.  
Fits were performed with $k_{\rm max}=8$, $\phi=1$, $t_0=0$, $|a_k|\le 10$. 
\label{fig:rcut}
 }
\end{center}
\end{figure}  

We can effectively raise the isovector threshold by including the $\pi\pi$ continuum 
explicitly, as constrained by $\pi\pi$ production and $\pi\pi \to N\bar{N}$ data: 
\be\label{eq:Gcut}
G^{(1)}_E(t) = G_{\rm cut}(t) + \sum_k a_k z^k(t,t_{\rm cut}=16 m_\pi^2, t_0) \,,
\ee
where $G_{\rm cut}(t)$ is generated by (\ref{eq:pipiNN}) for $4m_\pi^2 < t < 16 m_\pi^2$.  
For $|F_\pi(t)|$ we take the four $t$ values close to production threshold from \cite{Amendolia:1983di} 
($0.101$ to $0.178\,{\rm GeV}^2$), and twelve $t$ values from \cite{Achasov:2005rg} 
($0.185$ to $0.314\,{\rm GeV}^2$).   
The product of the remaining kinematic factor and $f^1_+$ from \cite{Hohler}
is interpolated to the appropriate $t$ value, and the integral computed as a discrete sum. 
Using coarser bin size (e.g. 8 instead of 16 bins) has no significant effect, indicating that
discretization error is small.   
Estimating the remaining coefficients by modeling the 
$\pi\pi$ continuum contribution for $16 m_\pi^2 \le t \le 40 m_\pi^2$ using (\ref{eq:fourier}) and 
(\ref{eq:pipiNN}) at $\phi=1$ and $t_0=0$
gives coefficients 
$a_1\approx -4.5$, $a_2 \approx 2.2$, $a_3\approx 2.1$.   Setting $|\sin(k\theta)|$ in 
(\ref{eq:fourier}) yields $|a_k|\lesssim 5.0$ for the remaining contribution of the $\pi\pi$ continuum 
in this model. 

We fit using the same proton and neutron data as in Section~\ref{sec:neutron}.
The resulting fit coefficients $a_{k=1,2,3}$ for $Q^2_{\rm max}=1\,{\rm GeV}^2$, $\phi=1$, $t_0=0$ and $k_{\rm max}=8$ are 
$-1.93(6)$, $-0.5^{+1.1}_{-1.3}$, $2\pm 7$ for the isocalar form factor;
and 
$-3.40^{+0.09}_{-0.10}$, $3.7^{+1.7}_{-1.3}$, $3^{+5}_{-10}$ for the isovector form factor.   
The sign and approximate magnitude of the first coefficients 
agree with the remaining $\pi\pi$ continuum model discussed above in the isovector case; 
and with the $\omega$ pole model discussed at the end of Section~\ref{sec:pipi} for the isoscalar case.  
The sizable contribution of the isovector $a_{k=1}$ in this scheme can be traced to the residual effects of 
the $\pi\pi$ continuum, including the $\rho$ peak, near the higher threshold.   With no loss 
of model-independence, we can replace $G_{\rm cut}(t)$ above with a new $G_{\rm cut}(t)$ generated 
by (\ref{eq:pipiNN}) for  $4m_\pi^2 < t < 40 m_\pi^2$, i.e., with the $\pi\pi$ continuum modeled to larger $t$.  
The value $t_{\rm cut}=16 m_\pi^2$ remains the same. 
We emphasize that this does not introduce a model dependence, as any discrepancy between $G_{\rm cut}(t)$ and the 
true $\pi\pi$ continuum is accounted for by parameters in the $z$ expansion. 
The resulting central value and errors on the charge radius are changed minimally by 
this modification.  The isoscalar coefficients are also not significantly changed, while
the isovector coefficients become $1.07(10)$, $1.6^{+1.6}_{-1.5}$, $1^{+7}_{-8}$.   
Figure~\ref{fig:rcut} shows the resulting extraction of the proton charge radius 
using for $G_{\rm cut}(t)$ the full model of the $\pi\pi$ continuum, and our default 
$\phi=1$, $t_0=0$.   As in Fig.~\ref{fig:Q2}, the inclusion of data beyond 
$Q^2 \sim {\rm few}\times 0.1\,{\rm GeV}^2$ has minimal impact on the fits.  

\section{Discussion \label{sec:discussion}} 

We have discussed determinations of the proton charge radius from 
the slope of the proton form factor $G_E^p(t)$, in four cases: 
(1) low-$Q^2$ electron-proton scattering data;  
(2) proton data including high $Q^2$; 
(3) proton plus neutron data; and (4) proton, neutron, and $\pi\pi$ data.  
We have investigated various expansion schemes, corresponding to choices of the parameter $t_0$ and
the function $\phi$, and shown that the impact on $r_E^p$ is minimal; in the following discussion we 
take $\phi=1$ and $t_0=0$. 

Including just the low $Q^2$ proton data~\cite{Rosenfelder:1999cd}, 
we find the result as in Table~\ref{tab:simplefits} [case (1)] 
$r_E^p = 0.877^{+0.031}_{-0.049} \pm 0.011 \,{\rm fm}$, where the first error 
is obtained using the more stringent bound $|a_k|\le 5$, and the additional error
is conservatively estimated by finding the maximum variation of the $\Delta \chi^2=1$ interval
when the fits are redone assuming $|a_k|\le 10$.   
Using a larger $Q^2$ range of proton data~\cite{Arrington:2007ux} decreases the uncertainty. 
Taking for definiteness $Q^2_{\rm max}=0.5\,{\rm GeV^2}$ and $k_{\rm max}=8$,  
we obtain via the same procedure, as in Fig.~\ref{fig:Q2} 
 [case (2)]
$r_E^p = 0.870 \pm 0.023 \pm 0.012\,{\rm fm}$. 
Including the neutron data, as in Table~\ref{tab:schemes}, we find  [case (3)] 
$r_E^p = 0.880^{+0.017}_{-0.020} \pm 0.007\,{\rm fm}$, where the same bounds, 
$|a_k|\le 5$, $|a_k|\le 10$ are enforced on both isoscalar and isovector coefficients and again $k_{\rm max}=8$.%
\footnote{
The slight difference between this value and that inferred from the final column for 
the first two rows of Table~\ref{tab:schemes} 
is due to the slight difference between $k_{\rm max}=6$ and $k_{\rm max}=8$.
}
The uncertainty induced by the 
neutron charge radius (\ref{eq:ren}) is negligible in comparison, $\lesssim 0.0005\,{\rm fm}$. 
Finally, including $G_{\rm cut}(t)$ as in (\ref{eq:Gcut}), we find [case (4)]
$r_E^p = 0.871 \pm 0.009 \pm 0.002 \pm 0.002 \,{\rm fm}$. 
For definiteness, we here include in $G_{\rm cut}(t)$ the extension of the 
$\pi\pi$ continuum model up to $t=40\,m_\pi^2$.  The first and the second error 
are as above, and the final error is obtained by assigning a $30\%$ normalization error 
to the continuum contribution, as discussed below.  

Let us compare our results to several previous determinations of $r_E^p$.   
Many of these suffer from model assumptions on the 
functional behavior of the form factor.   
The small uncertainties obtained by Simon et al.~\cite{Simon:1980hu} ($r_E^p=0.862\pm 0.012$) 
and by Rosenfelder~\cite{Rosenfelder:1999cd} ($r_E^p=0.880\pm 0.015$)
require inputs from higher $Q^2$ data, which however we do not believe were robustly estimated.     
We find that the coefficient of $t^2$ 
in the expansion of $G_E^p(t)$ [Eq.(\ref{eq:taylor})] is constrained by the Arrington et al. data 
compilation~\cite{Arrington:2007ux} to be 
$a_2^{\rm Taylor}/t_{\rm cut}^2 = 0.014^{+0.016}_{-0.013} \pm 0.005 \,{\rm fm}^4$ (using $Q^2_{\rm max}=1\,{\rm GeV}^2$, 
$k_{\rm max}=10$).   
A much smaller uncertainty, $a_2^{\rm Taylor}/t_{\rm cut}^2 = 0.011(4)\,{\rm fm}^4$ or $0.014(4)\,{\rm fm}^4$, 
was adopted in \cite{Rosenfelder:1999cd}.   
Even neglecting the additional uncertainty due to 
cubic and higher order terms, this would lead to a result 
$0.878\pm 0.008^{+0.047}_{-0.039}$ obtained using (\ref{eq:taylor}) and data as in Table~\ref{tab:simplefits}.  
The errors are from the data and from the first uncertainty 
on the quadratic coefficient, respectively.

The analyses of 
Sick~\cite{Sick:2003gm} ($r_E^p=0.895\pm 0.010\pm 0.013$) 
and 
Blunden and Sick~\cite{Blunden:2005jv} ($r_E^p=0.897\pm 0.018$) 
employ the continued fraction expansion (\ref{eq:cf}).   This functional form is unstable to 
the inclusion of additional parameters (see Table~\ref{tab:simplefits}), and error estimation relies 
on the investigation of model datasets.   
In this paper we have not fit directly to cross section data,
and we have not applied our analysis to this dataset. For a variation of this analysis see \cite{Borisyuk:2009mg}.

The dispersion analysis of Belushkin et al.~\cite{Belushkin:2006qa} 
($r_E^p=0.844^{+0.008}_{-0.004}\,{\rm fm}$, $r_E^p=0.830^{+0.005}_{-0.008}$) 
does not attempt to estimate uncertainties due to the constrained shape of
the assumed form factors.   
Our analysis makes clear which inputs have the most 
effect on the charge radius extractions.  In particular, data at large $|t|$, 
for either timelike or spacelike $t$, has minimal impact on fits to obtain 
$Q^2\approx 0$ quantities.   Inclusion of high-$Q^2$ data {\it does} 
introduce sensitivity to additional parameters, whose omission would introduce 
model dependence.   Our analysis provides a systematic procedure to analyze a wide
range of datasets in a model-independent way.   We emphasize that our goal is not 
simply reduction in the quoted error, but also the robust estimation of uncertainties.

Regarding the bounds on coefficients, in all approximations that we have considered
the bound $|a_k|\le 10$ appears very conservative.    
The sign and magnitudes of the first coefficients 
are consistent with expectations based on simple models, and it is rigorously true that
the coefficients $a_k$ must eventually decrease in magnitude for large $k$. 
At a practical level, the experimental 
determinations of these coefficients in each of the cases (1)-(4) above are consistent 
with magnitudes not larger than $|a_k| \sim 2$.  
Our implementation of the bounds on $a_k$ could be formalized in terms of standard 
methods of constrained curve fitting~\cite{Lepage:2001ym}.  
As discussed in ~\cite{Becher:2005bg}, our assumption of a flat ``prior'' should be conservative.  

Our analysis cannot discern inaccuracies in the datasets.  
For example, we have assumed that radiative corrections are properly accounted for in 
the compilations \cite{Rosenfelder:1999cd,Arrington:2007ux}, and that 
data correlations are sufficiently described by our treatment.%
\footnote{  
Two-photon exchange corrections were incorporated in \cite{Rosenfelder:1999cd}
using a simplified calculation of the Coulomb distortion.
See also \cite{Sick:2003gm,Blunden:2005jv}.   
Ref.~\cite{Arrington:2007ux} accounted for two-photon exchange
using the 
calculation of~\cite{Blunden:2003sp,Blunden:2005ew}.
}  
Within these assumptions, the values for cases (1)-(3) represent
model-independent determinations of the form-factor slope.    
Case (4) is more subtle.  While (\ref{eq:pipiNN}) is a model-independent relation
for the stated range $4m_\pi^2 \le t \le 16 m_\pi^2$, the determination of $f_+^1(t)$ 
in this range involves a dispersion relation with contributions from values of $t$ where the 
function is not rigorously constrained by continuation of $\pi N$ scattering data. 
Errors are not given in the tabulation \cite{Hohler}, 
and we are not aware of a critical assessment of the uncertainties associated with this analysis. 
It may be interesting to revisit this question. 
Ref.~\cite{Hohler:1974ht} suggests a $15\%$ error in the normalization of $f_+^1(t)$ at the $\rho$ peak; 
we take twice this value, $30\%$, as a representative uncertainty, which encompasses also 
the errors in $|F_\pi(t)|$.   
The resulting error for case (4) is thus not as rigorous, although the resulting 
$f_+^1(t)$ would need to be very different to become a dominant source of error. 

Within the stated uncertainties 
we find consistent results in each of our determinations, using both low and
high $Q^2$ proton data, neutron data, and pion continuum data.  
These methods can be applied to other datasets, and to fits using partial 
cross sections versus extracted form factors. 
For example, in a recent set of results~\cite{Bernauer:2010wm} the 
variation of $r_E^p$ under different model shapes for the form factor is larger than 
the other stated statistical and systematic errors. 
The same methods can be applied to other nucleon form factors and derived observables, 
including the axial-vector form factor probed in neutrino scattering~\cite{HillPaz}.   

\vskip 0.2in
\noindent
{\bf Acknowledgements}
\vskip 0.1in
\noindent
We have benefited from discussions with Z.-T.~Lu and C.~E.~M.~Wagner.   We also thank
J.~Arrington for discussions and comments on the manuscript. 
This work is supported by NSF Grant 0855039 and DOE grant DE-FG02-90ER40560.

\end{document}